\documentclass[conference]{IEEEconf}

\usepackage{graphicx} 
\usepackage{enumitem}
\usepackage{titlesec}
\input epsf
\usepackage{balance} 
\usepackage[table]{xcolor}
\usepackage{todonotes}
\usepackage{menukeys}
\usepackage{xcolor}
\usepackage{booktabs}
\usepackage{microtype}
\usepackage{hyperref}
\usepackage{multirow}
\usepackage{float}
\usepackage{booktabs}
\usepackage{array}
\usepackage{amsmath}
\usepackage{tcolorbox}
\usepackage{cite}
\usepackage{multirow}

\renewcommand\thesection{\arabic{section}} 

\renewcommand\thesubsectiondis{\thesection.\arabic{subsection}}

\renewcommand\thesubsubsectiondis{\thesubsectiondis.\arabic{subsubsection}}

\begin{document}

\title{\textbf{\Large Testing in the Evolving World of DL Systems: Insights from Python GitHub Projects\\}}

\author{Qurban Ali, Oliviero Riganelli, and Leonardo Mariani\\
	\normalsize University of Milano-Bicocca, Milan, Italy\\
	\normalsize q.ali@campus.unimib.it, oliviero.riganelli@unimib.it, leonardo.mariani@unimib.it\\

}

\maketitle
\begin{abstract}
In the ever-evolving field of Deep Learning (DL), ensuring project quality and reliability remains a crucial challenge. This research investigates testing practices within DL projects in GitHub. It quantifies the adoption of testing methodologies, focusing on aspects like test automation, the types of tests (e.g., unit, integration, and system), test suite growth rate, and evolution of testing practices across different project versions. We analyze a subset of 300 carefully selected repositories based on quantitative and qualitative criteria. This study reports insights on the prevalence of testing practices in DL projects within the open-source community.

\end{abstract}
\IEEEoverridecommandlockouts
\vspace{1.5ex}

\IEEEpeerreviewmaketitle


\section{Introduction}

The rapid growth of Deep Learning (DL) technologies has led to an increasing number of open-source projects incorporating DL components to meet requirements~\cite{klinger2018deep}. However, this trend poses challenges to software development practices~\cite{sarker2021deep}. Validating DL projects involves testing against components and models with limited control and transparency compared to traditional software components. As a result, understanding how practitioners validate DL projects and DL components becomes of crucial interest. 

In this study, we delve into the testing practices within DL projects on GitHub, focusing on aspects related to test implementation, test adoption, and test evolution. By analyzing 300 carefully selected repositories, we quantify the adoption of testing tools in the open-source DL community. We target Python as the primary language for our investigation due to its widespread use in DL application development~\cite{raschka2020machine}. By examining GitHub projects, we seek to understand how the open-source community approaches the validation of DL applications.

Through a systematic methodology involving quantitative and qualitative analyses, we formulate four research questions considering the presence and the types of tests implemented (RQ1), the adoption of test automation tools (RQ2), the consideration of test coverage metrics (RQ3), and the evolution of the test suites (RQ4).  

Our analysis of deep learning projects on GitHub revealed both promising trends and areas for improvement in testing practices. While a majority of projects incorporate test suites, only a portion focus specifically on the models themselves, suggesting a gap in model-specific validation. Functional testing dominates, with a focus on unit testing individual components.  The positive information is the significant adoption of test automation frameworks and GitHub workflows, but integrating automated test execution remains less common. Code coverage practices are moderate, with limited use for model-specific tests. The most notable finding is the intense test suite maintenance activity and rapid growth, highlighting the critical need for specialized methods for test case selection, prioritization, and regression testing tailored to the unique challenges of deep learning models. This research contributes to the understanding of testing practices in the context of DL projects, while opening further explorations in the evolving field of validation of DL software.

The rest of the paper is organized as follows. Section~\ref{sec:related} discusses related work. Section~\ref{sec:methodology} presents the research questions and the methodology used to answer them. Sections~\ref{sec:rq1}, \ref{sec:rq2} \ref{sec:rq3}, and \ref{sec:rq4} present the results for the four research questions investigated in this paper. Finally, Sections~\ref{sec:threat} and \ref{sec:conclusions} discuss threats to validity and provide final remarks, respectively.

\section{Related Work} \label{sec:related} 

The expanding presence of deep learning (DL) systems in safety-critical domains necessitates robust testing practices to mitigate faults and ensure reliable operation~\cite{humbatova2020taxonomy}. Humbatova et al. established a taxonomy of DL faults, highlighting the inherent complexity of these systems and the need for specialized testing approaches~\cite{humbatova2020taxonomy}. This gap in specialized testing motivates our investigation of testing practices within a large sample of open-source DL projects hosted on GitHub. We quantify the adoption and evolution of testing practices across different project versions, focusing on aspects such as test automation, the types of tests employed (e.g., unit, integration), test suite growth rate, and how these practices change over time. Several studies have explored software testing practices in general open-source projects~\cite{santos2023understanding,da2019adoption,ajila2007empirical,islam2023evolution,lin2020test}, providing a foundation for our research on DL projects.

Santos~\cite{santos2023understanding} presents an investigation of several Python machine-learning projects on GitHub, revealing that 65.19\% of them use at least one testing support library. They compared projects that utilized testing tools with those that did not, and findings reveal that projects employing automated testing tools exhibit lower instances of code smells, vulnerabilities, and bugs. The analysis of the tools revealed the prevalence of unittest, pytest, and doctest as popular choices for testing. Our study aligns with this investigation by systematically studying the adoption of testing tools in 300 Python DL open-source projects.

Silva \textit{et al.}~\cite{da2019adoption} investigate the adoption of automated support for testing in open-source projects across various programming language ecosystems. By analyzing 184 GitHub projects, the study reveals high adoption rates of testing in Go, PHP, and JavaScript, with specific mechanisms dominating each ecosystem, with adoption percentages ranging from 84.9\% to 100\% within the project corpus. An examination of test coverage was also conducted on 571 open-source projects. It revealed that the Python project, on average, exhibited a code coverage of 80.84\%. The authors also reported a Python test adoption rate of 37.5\%. Differently from this study, our investigation specifically targets DL projects, whose developers follow distinct practices and use different tools.

Kochhar \textit{et al.}~\cite{ajila2007empirical} investigate the adoption of software testing in over 20,000 non-trivial open-source projects from GitHub. The study explores the correlation between test cases and various project development characteristics, including project size, team size, number of bugs, bug reporters, and programming languages. By analyzing these relationships, the research provides insights into the testing practices, highlighting the importance of testing for project quality and bug reporting. The results indicate that 61.65\% of the analyzed open-source projects have test cases and that C++, C, and PHP have a higher number, on average, of test cases per project. Their results showed that, on average, Python projects have 41 test cases per project. Later, Islam et al~\cite{lin2020test} replicated the research questions by Kochhar~\cite {ajila2007empirical} regarding software testing in Java projects on GitHub from 2012 to 2021, aimed to validate the stability of previous findings and understand how software testing practices have evolved over the past decade. In contrast to our research, their scope encompasses a broader range of projects, whereas we concentrate solely on Python-based deep learning projects and investigate the usage of tools within source code. 

Finally, Lin \textit{et al.}~\cite{islam2023evolution} present a comprehensive study on test automation practices in open-source Android apps, focusing on over 12,000 projects from various app markets over a period of 5 years. The main objective was to investigate the adoption of test automation in non-trivial apps, with a particular emphasis on UI and unit tests. The study employed a rigorous methodology involving data collection from GitHub repositories, filtering criteria to select relevant apps, and a survey of app contributors to gain deeper insights. The results of the study revealed that only 8\% of the non-trivial open-source Android apps had automated tests, with a significant gap between developers' perceived testing practices and actual implementation. In comparison, we deal with Python deep learning projects, where we found much higher test automation, an average of 77.3\%. Unit testing was found to be more prevalent than UI testing in mobile applications, and we observed the same for DL projects. 

\section{Methodology} \label{sec:methodology}

This section outlines the research methodology presenting the research questions and the project selection criteria.

\subsection{Research Questions} \label{subsec:rq}
 To investigate testing practices within Python DL projects, we formulated the following research questions:
\begin{enumerate}[label=RQ\arabic*]
\item \textit{\textbf{Test Cases Presence}: Are Python DL GitHub projects tested? And what types of tests are implemented?}  This research question investigates the number and types of tests implemented to test DL Github projects.
To address this RQ, we combine quantitative and qualitative analyses. We first identify the testing frameworks and libraries used within the selected DL projects from the data extracted from the version control system (VCS). Additionally, we examine textual data sources, such as commit messages and code comments, to identify the types of tests implemented.

\item \textit{\textbf{Test Automation}: What is the adoption rate of test automation in Python DL GitHub projects?} This research question quantitatively investigates the usage of tools and libraries for test automation within Python DL GitHub projects. 

\item \textit{\textbf{Test Coverage}: Are test coverage criteria considered in Python DL GitHub projects?} 
Test coverage reports are valuable measures of the thoroughness of a test suite. This RQ investigates the adoption of coverage reports in Python DL projects, starting from the content of commit messages, code comments, and messages discussing test completeness. 

\item \textit{\textbf{Test Evolution}: How do test suites evolve in Python DL GitHub projects?} 
Measuring the evolution of test suite sizes in GitHub repositories can provide insights into the growth and performance of the tests over time. In particular, we consider the size and modification rate of test suites over time, based on the data present in VCS. 
\end{enumerate}

\subsection{Project Selection} ~\label{subsec:projectSelection} 
To ensure our findings effectively answer the research questions introduced in Section~\ref{subsec:rq}, we employed a multi-stage selection process to identify relevant Python deep learning (DL) projects hosted in GitHub. This process aimed to collect data from popular and actively maintained projects, with a specific focus on projects utilizing popular deep learning frameworks, such as TensorFlow~\cite{TF}, Keras~\cite{keras}, and PyTorch~\cite{PT}~\cite{yapici2021performance}~\cite{ibm}~\cite{viso}. These frameworks are widely adopted and reflect the ongoing trends in deep learning development~\cite{yapici2021performance}.

\subsubsection{Initial Search}

We leveraged the GitHub Search API~\cite{github} to retrieve a comprehensive set of projects for our analysis. Since we target Python DL projects, we used the term \texttt{\small language:python} in the search, combined with any of the three following learning frameworks \texttt{\small tensorflow}, \texttt{\small keras}, and \texttt{\small pytorch}.  As a result of our query, we retrieved: 50,500 projects using TensorFlow, 24,800 projects using Keras, and 64,000 projects using PyTorch. Since it would be infeasible to analyze such a large number of projects, we introduced project selection criteria for repository selection in the next section. 

\subsubsection{Project Selection Criteria}

We applied keywords and exclusion criteria to ensure we only chose relevant repositories for our study. For this purpose, we used the advanced search option available in the GitHub API to first \emph{exclude} personal repositories and projects with little adoption. That is, we included projects with five or more contributors by applying \textit{\texttt{"followers$\geq$ 5"}}; and repositories which are well received, with over 100 stars and 100 forks by using the \textit{\texttt{"stars > 100"}} and \textit{\texttt{"forks > 100"}} arguments, respectively. These metrics indicate the repository's impact and popularity because users typically star, watch, or fork repositories they find useful and interesting, thereby increasing their visibility within the GitHub community. This process resulted in 783 TensorFlow projects, 291 Keras projects, and 1200 PyTorch projects for further analysis.

Since the process of answering our research questions requires some manual analysis, we decided to limit the selection to 100 projects per DL framework. Before making our selection, we excluded tutorials, books, collections of code examples, or any projects that did not represent actual software systems used in practice by the developers. Furthermore, false positives projects that contain the specified search strings in the Python files but do not use the associated framework in real were also eliminated.  Repositories with minimal history, consisting of less than 100 commits, and inactive repositories that are no longer maintained were also excluded.

Instead of randomly selecting the projects, we finally selected the 300 repositories referring to quantifiable metrics. While there is no single metric that can fully capture a repository's value on GitHub, we choose to prioritize repositories with the highest star count. We finally selected the top 100 projects per framework according to this criterion. The minimum, average, and maximum number of stars, forks, and contributors for the selected projects are shown in Table~\ref{tab:number_of_stars_forks_contributors}. The material used to conduct our study is publicly available at~\cite{RP}.

\begin{table}[!tbp]
\centering
\caption {Distribution of Stars, Forks, and Contributors per Framework (Max/Average/Min)}
\resizebox{0.48\textwidth}{!}{%
\begin{tabular}{|l|c|c|c|c|c|c|c|c|c|}
\hline
\textbf{Framework} & \multicolumn{3}{c|}{Stars} & \multicolumn{3}{c|}{Forks} & \multicolumn{3}{c|}{Contributors} \\

\cline{2-10}
 & Max & Avg & Min & Max  & Avg & Min & Max & Avg & Min \\
\hline
TensorFlow & 115,000  & 121 & 8982  & 206,636  & 2066 & 103 & 2,183  & 16684 & 5\\
\hline
PyTorch & 115,000  & 11602 & 103 &  83,000  &  4550 & 62 & 10,900 & 299 & 7\\
\hline
Keras  & 27,300  &  4369 & 169 & 27,300  & 106621 & 42 &  843  &  80 & 5\\
\hline
\end{tabular}%
}
\label{tab:number_of_stars_forks_contributors}
\end{table} 

\section{RQ1: Test Case Presence} \label{sec:rq1} 
We begin our study by investigating if DL projects are tested, in particular, if the model itself is tested, and the types of tests that are implemented. 

To answer this question, we followed a systematic methodology consisting of three steps: (a) we determined if any test suite is present in the analyzed repository, (b) if present, we collect quantitative data about the existing test cases, and (c) we analyze the tests to determinate their types. 

\subsection{Determining the Presence of Test Suites} 
 
We first look for the presence of test suites and the type of testing being performed in the selected projects. To determine the percentage of projects that have test suites, we checked if they include a collection of test files/directories. We are however more interested in looking into model-related tests, that is, tests designed to exercise the model, or software features that use the model. The presence of model-related test files is important because these tests help to detect and prevent issues or bugs that may depend on the model. Their presence also indicates that the developers are committed to ensuring the code quality, reliability, and maintainability of the model.

It can be challenging to distinguish between model and non-model tests because there is not a clear demarcation between these two categories within the test suites. However, there are also easy cases, for instance when the codebase includes a directory for model files. To investigate the presence of test suites and model-specific tests within the projects, we adopted a multifaceted approach, exploring the project's documentation and utilizing keyword-based searches within the codebase. 

To identify tests and model-related tests, we analyzed the file structure within each repository to look for common naming conventions for test files that include \textit{\texttt{"test*.py"}} and \textit{\texttt{"*test.py"}} or directories named ‘tests’ or ‘test’ and for model-related tests that include but are not limited to ‘model\_test’, ‘test\_prediction’, ‘test\_training’, ‘test\_evaluation’, ‘test\_layers’, ‘test\_inference’, ‘test\_regression’, and ‘test\_classification’ within the test suite. We also look for the existence of the test-related frameworks or tool configuration files, including ‘pytest.ini’, ‘setup.cfg’, and ‘tox.ini’ which are normally used in testing setups. We also performed a targeted search using the specific keyword ‘test’ through GitHub's search option for each repository. This approach was employed particularly for projects where we could not initially identify the existence of a separate test suite. 

After completing this targeted search, we discovered that 232 out of 300 projects (77\%) contained test files (See Table~\ref{tab:test_types}). We also examined the project's documentation and README file to gain insights about the tests, but only five projects provided explicit descriptions of testing within their documentation.  
\subsection{Determining the Types of Tests} 
Since tests might be different in nature, validating different aspects of a developed project, our investigation takes the type of the implemented tests into account. In particular, we discriminate among the following types of tests.

\subsubsection{Functional Testing:}
\begin{itemize}
\item \textit{Unit Tests:}
Unit tests are tests used to test methods and functions of individual components or modules used by the software. These tests are closely tied to the source code and their execution is relatively economical to automate, using tools such as PyUnit, and UnitTest~\cite{sneha2017research}.

\item \textit{Integration Tests:} 
Integration tests are tests used to verify that multiple components or modules work well together. The primary objective here is to identify flaws within the interface, communication channels, and data exchange among software modules. Integration tests tend to be more resource-intensive than unit tests as they require the operation of multiple components~\cite{sneha2017research}. 

\item \textit{System Testing:} 
System tests emulate end-user interactions with an application in a real-world environment. They ensure the proper execution of various user tasks from simple actions, like logging in or loading a web page, to more complex tasks, such as validating email notifications and online payment processes. These tests are very useful but come with a higher cost in terms of maintenance and execution~\cite{sneha2017research}.

\end{itemize}
\subsubsection{Non-Functional Testing:}
\begin{itemize}
\item \textit{Performance Testing:}
Performance tests assess the system’s stability and response time under a specific workload. Additionally, it helps to measure the speed and scalability of the system~\cite{Wiley}. 

\item \textit{Smoke Testing:}
Smoke testing is a type of testing performed to ascertain the fundamental functionality of an application at a very high level. Their purpose is to provide an initial assurance that critical system features operate as expected. Smoke tests are useful for assessing the newly built software and ensuring that no major issue exists by verifying proper functionality following deployment to a fresh environment~\cite{Wiley,icst2023}.

\item \textit{Security Testing:}
Security testing is used to check if the application is secure from internal and external threats. It also checks how the application behaves against any attack or malicious programs~\cite{Wiley}.  

\end{itemize} 

To identify the test types, we specifically searched for the common naming conventions used for test folders/files within the test suite that includes ‘unit\_tests’, ‘integration\_tests’, ‘system\_tests/end\_to\_end\_tests’, ‘performance\_tests’, ‘smoke\_tests’, and ‘security\_tests’ for the unit, integration, system, performance, smoke, and security tests respectively, since normally developers name the tests, or the enclosing folders, consistently with their intended use. 

Additionally, we employed keyword searching to identify test files containing keywords that include ‘unittests’, ‘integration’, ‘system/end\_to\_end’, ‘performance’, ‘smoke’, and ‘security’. We also checked for the test files named according to the methods or components being tested, such as ‘Test\_MethodName.py’ or ‘Test\_FeatureName.py’.

Identifying integration tests can be tricky because developers do not always use a standard way to name them when organizing project files. Although, it can differ based on how developers organize their projects and the names they choose for integration files, such as ‘userservice\_integration\_test’ or ‘api\_integration\_test’, so we focused more on those files to correctly classify them as integration tests. 

We also examined the content of test files for scenarios that include interactions between various components or the entire system to classify them as an integration or system test, respectively.
For performance tests, we also searched for test files named \texttt{"load"}, \texttt{"stress"}, and \texttt{"concurrent\_user"} since these are used to measure the performance of the system. 

To mitigate any threat related to keyword-based searching, we carefully chose the set of keywords we used for searching, and more importantly, we manually verified the search outcome to mitigate any miss-classification risk. 

\subsection{Results}
Our analysis of the 300 projects yielded several key insights regarding test implementation. As shown in Table~\ref{tab:test_frameworks}, 77\% of the projects have a dedicated test suite. A non-trivial number of projects, 23\%, lacked any test suites, indicating the need to improve test adoption in Python DL projects.

The presence of model-specific tests, designed to validate model behavior and performance,  is indeed worse.  Only 55\% of the projects included model-specific tests. These results indicate an awareness of the need to test core model components but also highlight the need to improve testing practices, especially concerning model tests. These observations are quite consistent across the projects that use any of the three considered frameworks.

\begin{table}[!tbp]
\centering
\caption{{Test Occurrence}}
\resizebox{\columnwidth}{!}{%
\begin{tabular}{|l|c|c|c|c|c|}
\hline
\textbf{Framework} & \textbf{Projects with } & \textbf{No.of Untested} & \textbf{Proportion of Projects}  \\
 & \textbf{Test Suite}  & \textbf{Projects} & \textbf{with Model Tests} \\
\hline
{TensorFlow}  & 78 & 22 & 39/78 (50\%)   \\
{Keras} & 75 & 25 & 40/75 (53\%)  \\
{PyTorch} & 79 & 21 & 49/79 (62\%)  \\
\hline
\textbf{Total} & 232 (77\%) & 68 (23\%) & 128/232 (55\%) \\
\hline
\end{tabular}%
}

\label{tab:test_frameworks}
\end{table}

When considering the types of tests implemented within these repositories, our analysis revealed a predominant presence of functional tests, with a limited number of tests related to non-functional aspects, as reported in Table~\ref{tab:test_types}. 
Almost every project incorporates functional test cases, implemented as unit or integration tests, while system testing is less frequent.

Non-functional tests that assess aspects such as performance, security, and usability, were notably underrepresented within these projects in contrast to functional, as shown in Table~\ref{tab:test_types}. A small subset of projects, specifically 16 for TensorFlow, 10 for PyTorch, and 12 for Keras were classified as performance tests. Additionally, we found that 5 PyTorch and 3 Keras projects contain smoke tests. Furthermore, we did not find any evidence of security testing in Keras, while only 1 TensorFlow and 3 PyTorch projects contain security tests. This result indicates that developers of Python DL systems may need to pay more attention to the validation of the non-functional aspects. 

There could be several possible reasons behind this disparity which include the developer’s focus on core functionality, resource constraints, and complexity. Developers prioritize functional testing for its immediate value and ease of automation, which focuses on ensuring the software correctly performs its intended tasks ~\cite{yaseen2020prioritization}, while non-functional testing requires more resources, time, and expertise.  Non-functional aspects like performance and security can be more challenging to define metrics for and automate testing compared to functional testing ~\cite{camacho2016agile}. 

To address this, projects should integrate non-functional requirements early, invest in automated tools for non-functional tests, and continuously monitor to ensure balanced attention to both functional and non-functional aspects of testing, leading to more robust software.

\begin{table}[!tbp]
\centering
\caption{Type of test cases implemented in Python GitHub Projects}
\resizebox{0.49\textwidth}{!}{%
\begin{tabular}{|l|c|c|c|c|}
\hline
\textbf{Types of Test} & \multicolumn{4}{c|}{\textbf{DL Frameworks}} \\
\cline{2-5}
& \textbf{TensorFlow} & \textbf{PyTorch} & \textbf{Keras} & \textbf{Total} \\
\cline{1-5}
Unit & 39 & 54 & 41 & 110 (37\%) \\
Integration & 17 & 12 & 10 & 39 (13\%) \\
Smoke & - & 5 & 3 & 8 (3\%) \\
Performance & 16 & 10 & 12 & 28 (9\%) \\
System & 1 & - & 3 & 4 (1\%)\\
Security & 1 & 2 & - & 2 (1\%)\\
\hline
\end{tabular}%
}

\label{tab:test_types}
\end{table}

\begin{tcolorbox}
\emph{Answer to RQ1}: Most of the projects analyzed Python DL open-source projects include test cases, but only 55\% of them include tests specifically designed to exercise the models. Testing effort is largely focused on functional testing, with higher effort on unit testing, and smaller effort on integration and system testing. We recognized some attention to performance testing (9\% of the projects), and nearly missing evidence of smoke (3\% of the projects) and security (1\% of the projects) testing.
\end{tcolorbox}

\section{RQ2: Test Automation frameworks/libraries} \label{sec:rq2}

This question investigates the usage of automation tools, which play an important role in test case execution. In fact, manual testing is usually, expensive, prone to human errors, and hardly reproducible. On the contrary, automated test execution usually offers more robust, reliable, and repeatable testing processes, although the quality of the resulting process depends on the quality of the underlying test scripts and frameworks~\cite{leitner2007reconciling}. 

To determine if a GitHub test suite uses any automation tool to execute test cases automatically, we reviewed the repository's test files, code-base, and workflows to collect data about the automation tools used to execute the test cases. Although we selected Python projects, our analysis revealed that some projects also incorporate C++, C\#, Java, JavaScript, .Net, and Node.js code. We thus considered not only test automation tools for Python but also tools related to these languages to get a better picture of the automation tools used in the selected repositories. 

To identify the tools used in the test suites, we searched for specific folders according to common coding patterns, such as \textit{‘test/automation’}, \textit{‘tests/(language\_name)/automation’},  \textit{‘test/java/automation’} or \textit{‘test/python/automation’}, and naming conventions for test files, such as \textit{‘auto\_test*.py’} or \textit{‘auto\_test*.js’} which suggests the presence of test automation frameworks. We looked for the most commonly used frameworks for the following programming languages: \textit{PyTest~\cite{pytest}, Unittest~\cite{UnitTest}, JUnit~\cite{junit}, TestNG~\cite{testng}, Jest~\cite{jest}, Mocha\cite{mocha}, GTest\cite{googletest}, xUnit~\cite{xunit}} for \textit{Python, Java, JavaScript, Node.js, C++, and C\#}~\cite{AuT,da2019adoption}. This corresponds to searching for the keywords that identify the testing frameworks, which include, pytest, unittest, junit, testNG, jest, mocha, gtest, tox, and xunit.

We also thoroughly reviewed the project's documentation and README file, seeking any indication of test automation frameworks. During our investigation, we discovered that out of the several projects reviewed, only a small subset of them, specifically 20 out of 232 projects, provided additional information about their testing practices, supplementing our initial findings.

GitHub workflows are commonly used to test execution. Throughout our comprehensive analysis, we realized that more than half, precisely 161 out of 300 of the selected projects, contained workflow folders/files within their repositories. To extract additional information about testing practices, we carefully inspect the contents of the GitHub/workflows directory to identify any workflow files typically written in YAML format (i.e. ci.yml, travis.yml, and config.yml). These files serve as blueprints for defining the necessary steps to build, test, and deploy the codebase. Specifically, we focus on identifying workflow files that pertain to testing, as their presence indicates the utilization of an automated test suite. Apart from other workflow (yml) files some workflows contain specific automated tasks, such as installing test requirements, debugging CLI, running tests, and uploading coverage reports. Additionally, we also looked for Continuous Integration / Continuous Deployment (CI/CD) tools including TravisCI, Jenkins, CircleCI, and GitHub Actions which are sometimes used to automate testing. 

We inspected these workflow files to discover steps that involve executing test commands or employing testing frameworks. The inclusion of commands such as ‘npm test’, ‘run tests’, ‘pytest’, ‘pytest-cov’, ‘run tox’, ‘mocha’, or other similar testing commands serves as evidence of automated testing practices. These commands can be manually executed by developers or seamlessly integrated into continuous integration (CI) or continuous delivery (CD) pipelines to ensure consistent and automated testing throughout the development lifecycle.

\subsection{Results}

Among the projects under inspection, our analysis revealed that a substantial portion of these projects have adopted test automation. Automated testing adoption is identified in a project when it integrates at least one library related to test automation into its source code. Our analysis shows that 86.63\% of all projects adopt some form of test automation, to be precise 67 in the case of TensorFlow, 65 for Keras, and 69 for PyTorch. Table~\ref{tab:automation_frameworks} illustrates the distribution and percentages of Python DL projects implementing automated testing across various categories. 

\begin{table}[!tbp]
\centering
\caption{Test Automation Frameworks}
\resizebox{0.49\textwidth}{!}{%
\begin{tabular}{|l|c|c|c|c|c|c|c|c|c|}
\hline
\textbf{Automation Framework} & \multicolumn{5}{c|}{\textbf{DL Framework}} \\
\cline{2-6}
& \textbf{TensorFlow} & \textbf{PyTorch} & \textbf{Keras} & \textbf{Total} & \textbf{Percentage}\\
\cline{1-6}
{PyTest} & 34 & 54 & 41 & 129 & 48 \% \\

{Unittest} & 41 & 39 & 30 & 110 & 40 \% \\

{JUnit} & 2 & 4 & 4 & 10 & 3 \% \\ 

{Gtest} & 4 & 2 & 1 & 7 & 2.6 \% \\

{Jest} & 3 & 2 & 1 & 6 & 2.2 \% \\

{Tox} & 2 & 0 & 4 & 6 & 2.2 \% \\

{xUnit} & 1 & 0 & 0 & 1 & 0.37 \% \\ 

{TestNG} & 1 & 0 & 0 & 1 & 0.37 \% \\

{Mocha} & 0 & 0 & 0 & 0 & 0 \% \\


\hline
\end{tabular}%
}
\label{tab:automation_frameworks}
\end{table}

Results show that Pytest and Unittest are the most used automation frameworks with percentages of 48\% and 41\%, respectively. The presence of frameworks targeting other languages indicates that projects are often multi-language and require combining multiple tools to implement thorough test suites. 

Our investigation also reported that 53.6\% (161 projects) of the analyzed projects included GitHub workflow directories. Within these workflows, only one-third included test automation (54 projects). This result suggests that test automation is much more common than the definition of automated workflows that can execute automatic tests.

\begin{tcolorbox}
\emph{Answer to RQ2}: Unit test automation is quite common in Python DL open source projects, with 48\% of the projects using PyTest and 40\% of the projects using Unittest. Automated GitHub workflows are also quite common, with 53.6\% of the projects including them, but only a small fraction of the projects included automated test execution as part of the workflows.
\end{tcolorbox}

\section{RQ3: Code Coverage Analysis} \label{sec:rq3}

This research question delves into the adoption of test coverage practices within Python DL projects hosted in GitHub. Code coverage metrics are normally computed to measure the proportion of source code exercised by the test suite. It serves as a valuable metric to gauge the comprehensiveness of a project's test suite.  Common code coverage metrics include function, branch, statement, condition, and line coverage~\cite{10.1145/1138929.1138949}.

High code coverage increases confidence in the code since it demonstrates that a significant portion of the codebase has been actively exercised. This can help to reduce the risk of bugs escaping detection and impacting production systems~\cite{8031982}. Additionally, code coverage metrics can help identify areas of the code that require more testing attention.  Here, we investigate the extent to which these projects consider test coverage criteria, focusing on evidence found in commit messages, code comments, and discussions about test completeness.
 
To answer this RQ, we first examined project repositories to find the presence of test coverage reports. This includes examining commit messages, code comments, and discussions related to test completeness. We searched for comments that explicitly mention test coverage metrics (e.g., statement coverage, branch coverage), test coverage tools, or discussions about improving the thoroughness of the test suite.

Manually setting up and running coverage tools across numerous projects locally is both time-consuming and prone to errors. To streamline this process, we prioritized publicly available coverage information found in the project documentation. 
We began by thoroughly examining project READMEs. Ideally, READMEs provide labels denoting the employed coverage tools, offering easy access to comprehensive reports.
To strengthen our investigation and ensure we captured all potential test coverage practices, we conducted further examination beyond the initial README searches. This additional effort aimed to maximize our discovery of potentially used tools.
We examined documentation beyond the README files for mentions of code coverage tools. We inspected the ‘.github/workflows’ folder for files specifically associated with code coverage tools. We searched for filenames commonly associated with coverage tools, such as ‘.codecov.yml’, ‘.coveragerc’, ‘.coveralls.yml’, and ‘coverage.py’. For the projects with identified coverage tools, we analyzed the available reports to pinpoint areas with low coverage, particularly focusing on model-related files. 

\subsection{Results}
Our analysis revealed that 23\% (69 out of 300) of the projects included coverage badges within their documentation. These badges link to the Codecov~\cite{codecov}, Coveralls~\cite{coveralls}, and Scrutinizer~\cite{scrutinizer} tools, offering access to detailed code coverage reports (see Table~\ref{table:coverage_services}). Interestingly, while these badges point towards potential code coverage practices, a significant portion (approximately 77\%) of the projects lacked such badges, suggesting they might not be actively employing coverage tools.

Among the projects with coverage badges, Codecov emerged as the most popular choice, with 55 projects (18.33\%) utilizing it. Coveralls followed with 13 projects (4.33\%), and only one project (0.33\%) used Scrutinizer. It is worth noting that we also encountered 12 repositories containing files associated with coverage tools, but these lacked corresponding reports, potentially indicating abandoned or incomplete setups.

Focusing on the 69 projects with coverage reports, we observed that 19 (27.5\%) did not provide any details specifically related to model test coverage. This translates to 72.5\% of projects with reports including some level of coverage for model-related tests. However, a more concerning finding emerged when examining the broader pool of projects that included model-associated tests. Only 50 out of 128 such projects (39\%) had associated coverage reports. This significant discrepancy suggests potential shortcomings in testing practices for the core deep learning models themselves. Further investigation is necessary to understand the reasons behind this gap and assess the overall effectiveness of testing across these projects.

\begin{table}[!tbp]
\centering
\caption{Distribution of Code Coverage Tools}
\resizebox{0.7\columnwidth}{!}{%
\begin{tabular}{lcc}
\hline
\textbf{Coverage Tool} & \textbf{Count} & \textbf{Percentage} \\ \hline
Codecov & 55 & 18.33\% \\
Coveralls & 13 & 4.33\% \\
Scrutinizer & 1 & 0.34\% \\ 
\hline
\end{tabular} %
}

\label{table:coverage_services}
\end{table}

To analyze the depth of code coverage, we categorized projects based on their reported percentages (presented in Table~\ref{table:cov_categories}). The majority of projects (66.66\%) fell within the higher coverage categories (80-100\%). This distribution suggests that a significant portion of the projects target high coverage when coverage is computed. Conversely, few projects resided in the lower coverage categories. Only 5.80\% had coverage below 50\%, and 13.09\% fell within the 50-79\% range.

\begin{table}[!tbp]
\centering
\caption{Distribution of Projects With Code Coverage Percentage}
\resizebox{0.6\columnwidth}{!}{%
\begin{tabular}{lcc}
\hline
\textbf{Category} & \textbf{Count} & \textbf{Percentage} \\ \hline
Below 50 & 4 & 5.80\% \\
50-59 & 5 & 7.25\% \\
60-69 & 3 & 4.34\% \\ 
70-79 & 11 & 15.94\% \\
80-89 & 22 & 31.88\% \\
90-100 & 24 & 34.78\% \\ 
\hline
\end{tabular}%
}
\label{table:cov_categories}
\end{table}

\begin{tcolorbox}
\emph{Answer to RQ3}: Our analysis of test coverage practices in deep learning projects revealed moderate adoption of code coverage tools (23\% of the projects), with Codecov being the most popular. While a positive trend emerged with many projects achieving high coverage (56.66\% of the projects collecting coverage metrics report a coverage higher than 80\%), a significant portion lacked reports (77\% of the projects). More concerning, only 39\% of projects with model tests had associated coverage reports, suggesting potential shortcomings in testing core deep learning models. These findings highlight a need for further investigation into testing practices specifically tailored to deep learning models.
\end{tcolorbox}

\section{RQ4: Test suite growth and maintenance} \label{sec:rq4}

Studying the evolution of test suites in GitHub repositories provides valuable insights into the maintenance of test suites. Analyzing trends in the changes to test suites helps us pinpoint areas of code that have undergone substantial modifications, enabling a comprehensive assessment of software reliability. A consistently growing test suite often reflects a strong commitment to maintaining code quality and preventing regressions, ultimately contributing to increased software stability and reliability.

This examination includes observing the frequency of changes in the test suite, specifically concerning modifications to models or software components. Furthermore, the identification of changes, spanning additions, deletions, and modifications to test files, adds depth to our investigation. Our analysis reveals how changes in a test suite are intertwined with the overall software development process.

To answer the research question, we selected 15 projects whose evolution has been analyzed in detail. For these projects, we investigated the test suite growth by counting the frequency of test file addition, modification, and removal through version history, and the frequency of Pull Requests (PRs) \& commits related to testing within each repository.

The selection of the project has been driven by specific criteria to ensure the robustness and reliability of our analysis. To this end, we selected 15 repositories in total, selecting 5 for each framework (TensorFlow, Keras, and PyTorch).
To determine the projects, we selected the ones with the highest number of stars and forks, and those with model-related tests.  Additionally, we prioritized repositories that provided a coverage report, ensuring a more comprehensive and insightful analysis for our research. By employing these criteria, we aimed to narrow down our focus to repositories that not only represent the popularity of a framework but also exhibit a strong testing infrastructure, contributing to the validity and depth of our research findings. Table~\ref{tab:github_stats} lists the selected projects with their respective parameters, including contributors, stars, forks, and coverage reports.

\begin{table*}[!tbp]
\centering
\caption{List of the selected GitHub Repositories}
\resizebox{0.9\textwidth}{!}{%
\begin{tabular}{|l|l|c|c|c|c|c|}
\hline
\textbf{Framework} & \textbf{Project} & \textbf{Contributors} & \textbf{Stars} & \textbf{Forks} & \textbf{Coverage Report} \\
\hline
\multirow{5}{*}{\textbf{TensorFlow}} & \url{https://github.com/GPflow/GPflow} & 77 & 1800 & 436 & Yes \\
& \url{https://github.com/conda/conda} & 424 & 5900 & 1400 & Yes \\
& \url{https://github.com/deepchem/deepchem} & 161 & 4600 & 1500 & Yes \\
& \url{https://github.com/XanaduAI/strawberryfields} & 48 & 725 & 186 & Yes \\
& \url{https://github.com/kubeflow/pipelines} & 28 & 3200 & 1400 & Yes \\
\hline
\multirow{5}{*}{\textbf{PyTorch}} & \url{ttps://github.com/snorkel-team/snorkel} & 71 & 5600 & 856 & Yes \\
& \url{https://github.com/open-mmlab/mmdetection} & 436 & 26000 & 9000 & Yes \\
& \url{https://github.com/Lightning-AI/lightning} & 842 & 25100 & 3100 & Yes \\
& \url{https://github.com/pyg-team/pytorch_geometric} & 448 & 18900 & 3400 & Yes \\
& \url{https://github.com/arraiyopensource/torchgeometry} & 240 & 8700 & 886 & Yes \\
\hline
\multirow{5}{*}{\textbf{Keras}} & \url{https://github.com/IBM/adversarial-robustness-toolbox} & 102 & 4000 & 1100 & Yes \\
& \url{https://github.com/dask/distributed} & 310 & 1500 & 704 & Yes \\
& \url{https://github.com/wandb/wandb} & 159 & 7500 & 561 & Yes \\
& \url{https://github.com/scikit-learn-contrib/imbalanced-learn} & 78 & 6600 & 1300 & Yes \\
& \url{https://github.com/stellargraph/stellargraph} & 32 & 2800 & 419 & Yes \\
\hline
\end{tabular}%
}
\label{tab:github_stats}
\end{table*}

We start our investigation of test suite evolution from the commit history and pull requests, computing the percentage of commits related to tests. We leverage the GitHub Search API to identify commits and PRs related to testing. We search for commits and PRs containing the keyword 'testing' to retrieve relevant results. These indicators capture the degree of activity on the test cases, compared to the overall project activity.

To access commit messages, we used GitHub's API to extract the complete version control history for each selected repository. To ensure a representative sample of releases for our analysis, we strategically selected releases with uniform spacing between them, taking into account the total number of releases available. of each repository's release history. This approach maintained consistent intervals between the chosen releases. 

Furthermore, to ensure consistent comparisons across repositories with varying release history lengths, we analyze 10 releases from each project. 
To ensure a representative sample of releases for our analysis, we strategically selected releases with uniform spacing between them, taking into account the total number of releases available. of each repository's release history. This approach maintained consistent intervals between the chosen releases.

From the extracted data, we tracked the addition, modification, or deletion of test files and calculated the size of the test suite for each selected project. Our purpose is to assess the overall growth of test suites within the selected repositories over time. 
With this analysis, we computed the total number of test files per version and the percentage of test files modified in each version. 
   
When available, we finally tracked the code coverage metrics for each version, to evaluate the impact of test suite growth on the thoroughness of code testing. Correlating test suite growth with code coverage trends would help to determine whether the increase in test cases is effectively improving or keeping steady code coverage.

\subsection{Results}

\begin{figure}[!tbp]
\centering
\includegraphics[width=0.49\textwidth]{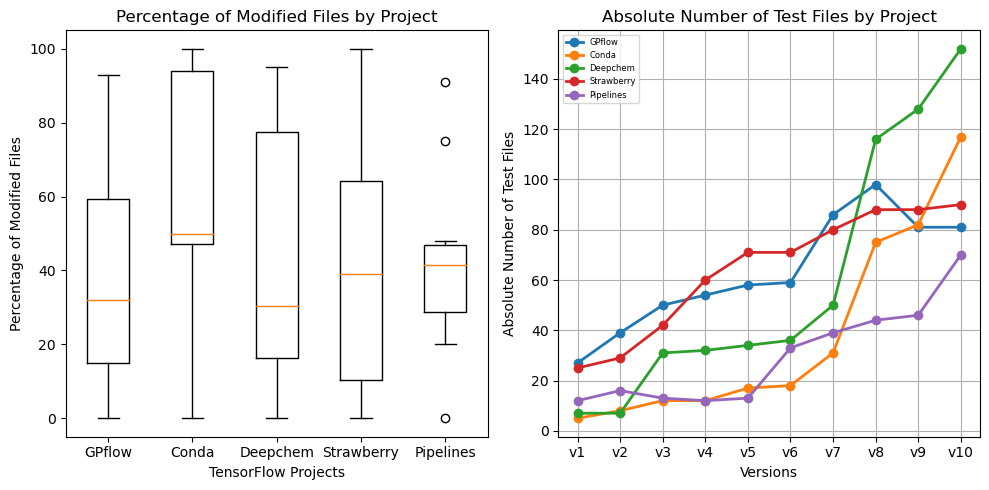}
\vspace{-8mm}
\caption{\label{fig:Tensor_M}TensorFlow Test Suite Growth.}
\end{figure}

\begin{figure}[!tbp]
\centering
\includegraphics[width=0.49\textwidth]{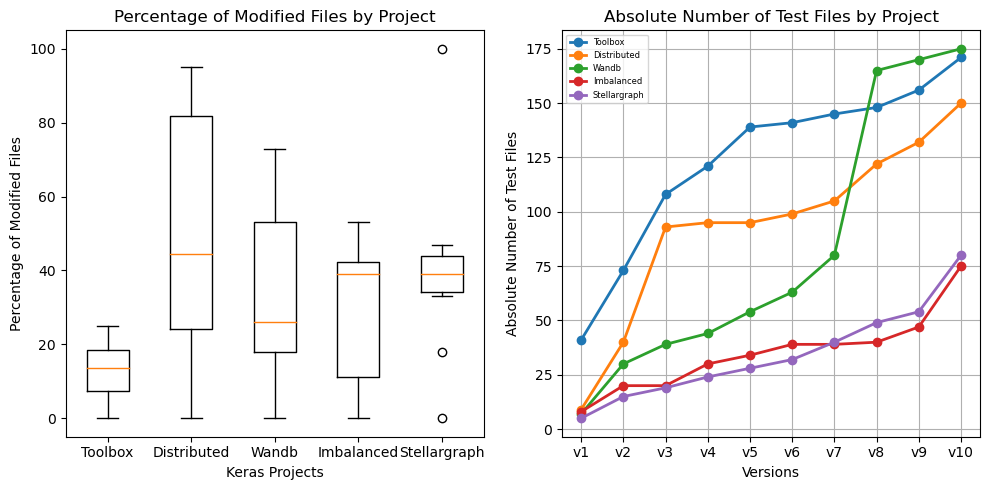}
\vspace{-8mm}
\caption{\label{fig:Keras_M}Keras Test Suite Growth.}
\end{figure}

\begin{figure}[!tbp]
\centering
\includegraphics[width=0.49\textwidth]{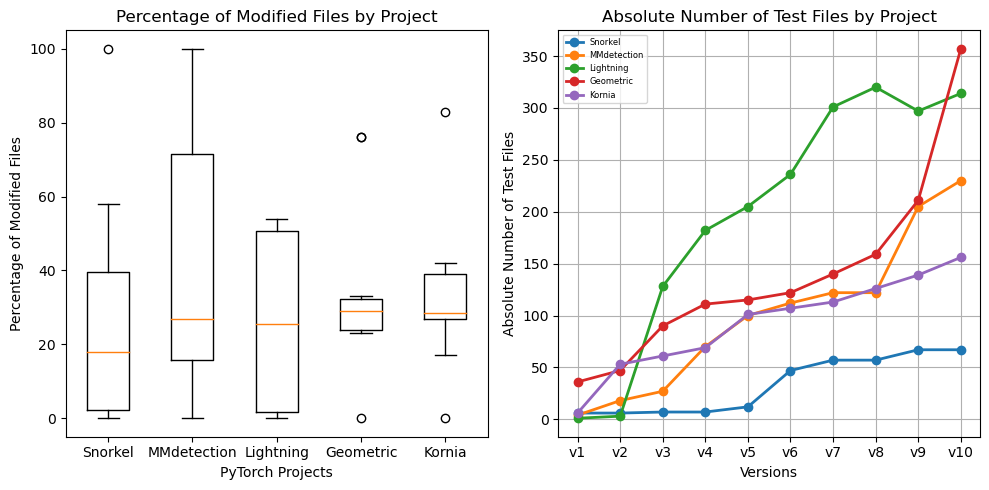}
\vspace{-8mm}
\caption{\label{fig:Pytorch_M}PyTorch Test Suite Growth.}
\end{figure}

In this section, we present the results from our investigation of the growth rate and maintenance of the test suite by comparing the changes in the test suite among the different versions of the selected repositories separately for each framework. Figures~\ref{fig:Tensor_M}, \ref{fig:Keras_M}, and \ref{fig:Pytorch_M} show the results for TensorFlow, Keras, and PyTorch for test suite modification and test suite growth, respectively. Test suite modifications show the percentage of test files changed per release, while test suite growth shows the total number of test files per project.

In terms of the range of changes per version, TensorFlow projects show a higher number of changes per test file, with changes that may reach 100\% of the files. This is not the case for Keras and Pytorch, where the amount of changes per version is typically smaller.  The different amount of changes also reflect the mean number of changes per version: it ranges between 35\% and 45\% for TensorFlow projects, while it stays between 15\% and 40\% for Keras projects, and between 20\% and 25\% for PyTorch projects. One of the possible reasons for the higher frequency of changes in TensorFlow projects may be its lower total count of test files compared to other frameworks. Further investigation into the nature of the changes, project types, practices, and testing culture within each framework is necessary to obtain a clearer picture.

Despite the different amounts of changes accumulated per version, all the projects show an overall increase in the number of tests per version, with some projects showing a rapid increase in the number of test files at certain points of their developments. Sometimes the number of test files grows as much as a factor of 10x in 10 releases. The pace of release cycles can significantly impact this growth in test files. While the exact cause for this spike is unclear, possible explanations could be the restructuring of the codebase, the introduction of new features in the project, or an increased emphasis on test coverage by the developers. This observation raises the attention toward regression, prioritization, and selection methods designed for DL projects, which may help cost-effectively manage test evolution.

Plots suggest there is a natural relation between the total number of test files per project and the number of test files modified across versions in TensorFlow projects. Projects with a higher number of total test files also tend to have a higher number of modified test files.

It is interesting to note that the project with the most overall test files (Toolbox) does not necessarily have the most significant modifications between versions. On the other hand, projects with the least number of overall test files, like Stellargraph, can have a significant increase in the percentage of test files modified. This suggests that the number of test files is not the only factor affecting the modifications made to the test suite.

\begin{table*}[!tbt]
\centering
\caption {DL Framework Commit Analysis}
\resizebox{0.9\textwidth}{!}{%
\begin{tabular}{|l|c|c|c|c|c|c|c|c|c|}
\hline
{\textbf{Framework}} & \textbf{Projects} & \textbf{Total No.} & \textbf{Testing} & \textbf{Per \%} & \textbf{Total No.} & \textbf{Testing} & \textbf{Per \%} & \textbf{Avg Testing} & \textbf{Avg Testing}\\
& & \textbf{commits} & \textbf{commits} & & \textbf{PRs} & \textbf{PRs} & & \textbf{Commit} & \textbf{PRs}\\
\hline
\cline{1-8} 
\multirow{5}{*}{\textbf{TensorFlow}} & 
{GPflow} & 2445 & 542 & 22\% & 1259 & 542 & 43\% & & \\
& {conda} & 16680 & 2000 & 12\% & 4036 & 1786 & 44\% & & \\
& {deepchem} & 14916 & 3000 & 20\% & 14410 & 8000 & 56\% & 25\% & 70.1\%\\
& {strawberryfields} & 1171 & 318 & 28\% & 607 & 314 & 52\% & & \\
& {pipelines} & 5679 & 1000 & 18\% & 6827 & 5580 & 82\% & & \\
\hline
\multirow{5}{*}{\textbf{PyTorch}} & 
{snorkel} & 2690 & 196 & 7\% & 742 & 395 & 53\% & & \\
& {mmdetection} & 2701 & 312 & 12\% & 3095 & 1515 & 49\% & & \\
& {lightning} & 10084 & 1000  & 10\% & 9861 & 8659 & 88\% & 11.7\%& 72.7\%\\
& {pytorch\_geometric} & 7374 & 880 & 23\% & 2820 & 1523 & 54\% & & \\
& {kornia} & 2642 & 598 & 23\% & 1747 & 1193 & 68\% & & \\
\hline
\multirow{5}{*}{\textbf{Keras}} & 
{toolbox} & 12141 & 1000 & 8\% & 1310 & 898 & 69\% & & \\
& {distributed} & 5670 & 996 & 18\% & 4900 & 3135 & 64\% & & \\
& {wandb} & 6073 & 858 & 14\% & 3954 & 3062 & 78\% & 13.2\% & 66.3\%\\
& {imbalanced-learn} & 855 & 72 & 9\% & 491 & 191 & 39\% & & \\
& {stellargraph} & 2535 & 378 & 15\% & 932 & 397 & 43\% & & \\
\hline
\end{tabular}%
}
\label{tab:commit_stats}
\end{table*}

Table~\ref{tab:commit_stats} presents the results from Commit and PRs analysis, where we aim to reveal the discussion around the testing practice among the developer community. The results show that TensorFlow has the highest average of testing commits accounting for 25\% while PyTorch has the lowest average accounting for 11.7\%. But surprisingly PyTorch has the highest average of PRs related to testing accounting for 72.7\% and Keras accounts for 66.3\% which is the lowest among the three frameworks.  Yet, the large majority of PRs in all the projects include discussion of both code and test changes.

Additionally, all the frameworks show varying percentages of testing commits and PRs across different projects. TensorFlow projects like GPflow and Conda have relatively lower testing percentages, and projects like Deepchem and Pipelines show higher testing percentages. The possible reason for this variation could be the nature of the projects and the emphasis on testing within their respective communities. PyTorch exhibits a similar trend, with projects like Geometric and Kornia showing higher testing percentages in comparison to others. However, Keras projects have moderate testing percentages in general, with projects like Wandb showing relatively higher testing percentages. The variation in testing percentages could be due to factors such as community size, project maturity, and the availability of testing tools and frameworks tailored for each framework.

\begin{tcolorbox}
\emph{Answer to RQ4:} We reported intense maintenance activity in the projects analyzed, with test suites that are extensively modified at every revision. Moreover, we reported a relevant increase in the number of tests per version, suggesting that tests are quickly accumulated in Python open-source DL projects, and the design of test case selection, prioritization, and regression methods that take the presence of models into account might be particularly relevant in this context.
\end{tcolorbox}

\section{Threats to Validity} \label{sec:threat}
This research is subject to limitations that could potentially influence the generalizability and conclusions drawn from the analysis of test suite growth and maintenance practices. Recognizing these threats to validity strengthens the transparency and trustworthiness of our findings.

\textit{\textbf{Selection Bias}}: The chosen sample of repositories might not fully represent the broader software development landscape. Focusing on popular repositories with high star and fork counts could skew the results towards projects with a stronger emphasis on testing practices. This may not reflect the practices of smaller or less popular projects. We attempted to mitigate this threat by employing specific criteria for repository selection and ensuring diversity in terms of framework, size, and domain.

\textit{\textbf{Data Extraction Limitations}}: Our reliance on commit messages and code comments for identifying test-related activities might overlook instances where developers do not explicitly mention testing in their commits. This could lead to an underestimation of the actual frequency of test suite changes. To mitigate this, we could explore advanced techniques for natural language processing to identify test-related activities within commits, even when not explicitly mentioned.

\textit{\textbf{Generalizability}}: The findings from the selected frameworks (TensorFlow, Keras, PyTorch) may not be generalizable to other programming languages or frameworks. The testing practices and culture could vary significantly across different software development communities. We attempted to mitigate this threat by clearly defining the scope and limitations of our study. Future studies could explore test suite growth patterns across a broader range of programming languages and frameworks to enhance generalizability.

\textit{\textbf{Metrics Limitations}}: The metrics used to measure test suite growth may introduce validity threats due to their subjectivity and potential biases. For example, counting the number of test-related comments or pull requests may not accurately reflect the actual level of testing activity within a repository. We attempted to mitigate this threat by using multiple metrics and cross-referencing data from different sources to validate our findings. 

\section{Conclusion} \label{sec:conclusions}
 Our investigation of Python deep learning projects on GitHub yielded valuable insights. Over 65\% of the analyzed projects incorporated some level of testing, demonstrating a growing trend of adopting testing practices in this domain. Interestingly, a diverse range of testing frameworks and libraries were employed across projects, highlighting the absence of a single standard approach.  Furthermore, the study revealed active maintenance of test suites, with projects tending to accumulate tests over time.  However, variations in testing practices were observed between projects utilizing TensorFlow, Keras, and PyTorch frameworks. 
These findings highlight the increasing significance of testing in the development of deep learning systems. Nevertheless, the research identifies the need for further exploration, particularly in the area of developing efficient strategies for managing and prioritizing tests specifically designed for deep learning models. Ultimately, this work emphasizes the importance of continued research to establish best practices and enhance the effectiveness of testing within deep learning projects.

\section*{Acknowledgements} 
 This work has been partially supported by the Centro Nazionale HPC, Big Data e Quantum Computing (PNRR CN1 spoke 9 Digital Society \& Smart Cities); the Engineered MachinE Learning-intensive IoT systems (EMELIOT) national research project, which has been funded by the MUR under the PRIN 2020 program (Contract 2020W3A5FY); and by the ReGAInS project funded by the MUR under the “Dipartimenti di Eccellenza 2023-2027" program.

\balance 

\bibliographystyle{IEEEtran}
\bibliography{references}

\begin{thebibliography}{10}
\providecommand{\url}[1]{#1}
\csname url@samestyle\endcsname
\providecommand{\newblock}{\relax}
\providecommand{\bibinfo}[2]{#2}
\providecommand{\BIBentrySTDinterwordspacing}{\spaceskip=0pt\relax}
\providecommand{\BIBentryALTinterwordstretchfactor}{4}
\providecommand{\BIBentryALTinterwordspacing}{\spaceskip=\fontdimen2\font plus
\BIBentryALTinterwordstretchfactor\fontdimen3\font minus \fontdimen4\font\relax}
\providecommand{\BIBforeignlanguage}[2]{{%
\expandafter\ifx\csname l@#1\endcsname\relax
\typeout{** WARNING: IEEEtran.bst: No hyphenation pattern has been}%
\typeout{** loaded for the language `#1'. Using the pattern for}%
\typeout{** the default language instead.}%
\else
\language=\csname l@#1\endcsname
\fi
#2}}
\providecommand{\BIBdecl}{\relax}
\BIBdecl

\bibitem{klinger2018deep}
J.~Klinger, J.~Mateos-Garcia, and K.~Stathoulopoulos, ``Deep learning, deep change? mapping the development of the artificial intelligence general purpose technology,'' \emph{arXiv preprint arXiv:1808.06355}, 2018.

\bibitem{sarker2021deep}
I.~H. Sarker, ``Deep learning: a comprehensive overview on techniques, taxonomy, applications and research directions,'' \emph{SN Computer Science}, vol.~2, no.~6, p. 420, 2021.

\bibitem{raschka2020machine}
S.~Raschka, J.~Patterson, and C.~Nolet, ``Machine learning in python: Main developments and technology trends in data science, machine learning, and artificial intelligence,'' \emph{Information}, vol.~11, no.~4, p. 193, 2020.

\bibitem{humbatova2020taxonomy}
N.~Humbatova, G.~Jahangirova, G.~Bavota, V.~Riccio, A.~Stocco, and P.~Tonella, ``Taxonomy of real faults in deep learning systems,'' in \emph{Proceedings of the ACM/IEEE 42nd international conference on software engineering}, 2020, pp. 1110--1121.

\bibitem{santos2023understanding}
\BIBentryALTinterwordspacing
W.~M.~A. dos Santos, ``Understanding the testing culture of machine learning projects on github.'' 2023. [Online]. Available: \url{http://dspace.sti.ufcg.edu.br:8080/xmlui/handle/riufcg/29359}
\BIBentrySTDinterwordspacing

\bibitem{da2019adoption}
R.~M. da~Silva, C.~Cruz, H.~de~S.~Campos, L.~G. Murta, and V.~de~Oliveira~Neves, ``What is the adoption level of automated support for testing in open-source ecosystems?'' in \emph{Proceedings of the IV Brazilian Symposium on Systematic and Automated Software Testing}, 2019, pp. 80--89.

\bibitem{ajila2007empirical}
S.~A. Ajila and D.~Wu, ``Empirical study of the effects of open source adoption on software development economics,'' \emph{Journal of Systems and Software}, vol.~80, no.~9, pp. 1517--1529, 2007.

\bibitem{islam2023evolution}
A.~Islam, N.~T. Hewage, A.~A. Bangash, and A.~Hindle, ``Evolution of the practice of software testing in java projects,'' in \emph{2023 IEEE/ACM 20th International Conference on Mining Software Repositories (MSR)}.\hskip 1em plus 0.5em minus 0.4em\relax IEEE, 2023, pp. 367--371.

\bibitem{lin2020test}
J.-W. Lin, N.~Salehnamadi, and S.~Malek, ``Test automation in open-source android apps: A large-scale empirical study,'' in \emph{Proceedings of the 35th IEEE/ACM International Conference on Automated Software Engineering}, 2020, pp. 1078--1089.

\bibitem{TF}
T.~G.~B. team, ``Tensorflow,'' \url{https://www.tensorflow.org/}, 2024.

\bibitem{keras}
F.~Chollet, ``Keras,'' \url{https://keras.io/}, 2024.

\bibitem{PT}
F.~A. research group, ``Pytorch,'' \url{https://pytorch.org/}, 2024.

\bibitem{yapici2021performance}
M.~M. YAPICI and N.~Topalo{\u{g}}lu, ``Performance comparison of deep learning frameworks,'' \emph{Computers and Informatics}, vol.~1, no.~1, pp. 1--11, 2021.

\bibitem{ibm}
\BIBentryALTinterwordspacing
S.~Madhavan. Compare deep learning frameworks. [Online]. Available: \url{https://developer.ibm.com/articles/compare-deep-learning-frameworks}
\BIBentrySTDinterwordspacing

\bibitem{viso}
\BIBentryALTinterwordspacing
V.~Meel. Top 10 deep learning frameworks in 2024. [Online]. Available: \url{https://viso.ai/deep-learning/deep-learning-frameworks/}
\BIBentrySTDinterwordspacing

\bibitem{github}
T.~Preston-Werner, ``Github search api,'' \url{https://docs.github.com/en/rest/quickstart?apiVersion=2022-11-28}, 2024.

\bibitem{RP}
L.~M. Qurban~Ali, Oliviero~Riganelli, ``Material,'' \url{https://github.com/lakhanqurban/PDLTesting}, 2024.

\bibitem{sneha2017research}
K.~Sneha and G.~M. Malle, ``Research on software testing techniques and software automation testing tools,'' in \emph{2017 international conference on energy, communication, data analytics and soft computing (ICECDS)}.\hskip 1em plus 0.5em minus 0.4em\relax IEEE, 2017, pp. 77--81.

\bibitem{Wiley}
\emph{The Art of Software Testing}.\hskip 1em plus 0.5em minus 0.4em\relax John Wiley \& Sons, Ltd, 2012, ch. 5,6,7.

\bibitem{icst2023}
C.~Cannavacciuolo and L.~Mariani, ``Smoke testing of cloud systems,'' in \emph{Proceedings of the International Conference on Software Testing, Verification and Validation}, 2022, pp. 47--57.

\bibitem{yaseen2020prioritization}
M.~Yaseen, A.~Mustapha, and N.~Ibrahim, ``Prioritization of software functional requirements from developers perspective,'' \emph{International Journal of Advanced Computer Science and Applications}, vol.~11, no.~9, 2020.

\bibitem{camacho2016agile}
C.~R. Camacho, S.~Marczak, and D.~S. Cruzes, ``Agile team members perceptions on non-functional testing: influencing factors from an empirical study,'' in \emph{2016 11th international conference on availability, reliability and security (ARES)}.\hskip 1em plus 0.5em minus 0.4em\relax IEEE, 2016, pp. 582--589.

\bibitem{leitner2007reconciling}
A.~Leitner, I.~Ciupa, B.~Meyer, and M.~Howard, ``Reconciling manual and automated testing: The autotest experience,'' in \emph{2007 40th Annual Hawaii International Conference on System Sciences (HICSS'07)}.\hskip 1em plus 0.5em minus 0.4em\relax IEEE, 2007, pp. 261a--261a.

\bibitem{pytest}
PyPI, ``Pytest,'' \url{https://docs.pytest.org/en/8.0.x/}, 2024.

\bibitem{UnitTest}
K.~Beck, ``unittest,'' \url{https://docs.python.org/3/library/unittest.html}, 2024.

\bibitem{junit}
------, ``Junit,'' \url{https://junit.org/junit5/}, 2024.

\bibitem{testng}
C.~Beust, ``Testng,'' \url{https://testng.org//}, 2024.

\bibitem{jest}
C.~Nakazawa, ``Jest,'' \url{https://jestjs.io/}, 2024.

\bibitem{mocha}
Eich, ``Mocha,'' \url{https://mochajs.org/}, 2024.

\bibitem{googletest}
I.~Google, ``Googletest,'' \url{https://github.com/google/googletest}, 2024.

\bibitem{xunit}
S.~Madhavan, ``Xunit,'' \url{https://xunit.net}, 2024.

\bibitem{AuT}
\BIBentryALTinterwordspacing
M.~Poliarush, ``awesome-test-automation.''\hskip 1em plus 0.5em minus 0.4em\relax GitHub, 2015. [Online]. Available: \url{https://github.com/atinfo/awesome-test-automation/}
\BIBentrySTDinterwordspacing

\bibitem{10.1145/1138929.1138949}
\BIBentryALTinterwordspacing
Q.~Yang, J.~J. Li, and D.~Weiss, ``A survey of coverage based testing tools,'' ser. AST '06.\hskip 1em plus 0.5em minus 0.4em\relax New York, NY, USA: Association for Computing Machinery, 2006, p. 99–103. [Online]. Available: \url{https://doi.org/10.1145/1138929.1138949}
\BIBentrySTDinterwordspacing

\bibitem{8031982}
P.~S. Kochhar, D.~Lo, J.~Lawall, and N.~Nagappan, ``Code coverage and postrelease defects: A large-scale study on open source projects,'' \emph{IEEE Transactions on Reliability}, vol.~66, no.~4, pp. 1213--1228, 2017.

\bibitem{codecov}
J.~Engelberg and E.~Hooten, ``Codecov,'' \url{https://about.codecov.io/}, 2024.

\bibitem{coveralls}
I.~Coveralls, ``Coveralls,'' \url{https://coveralls.io/}, year={2024}.

\bibitem{scrutinizer}
N.~C. Zakas, ``Scrutinizer,'' \url{https://scrutinizer-ci.com/}, 2024.

\end{thebibliography}

\balance

\end{document}